\documentclass[preprint]{aastex}
\def\beq{\begin{equation}}
\def\eeq{\end{equation}}
\def\bey{\begin{eqnarray}}
\def\eey{\end{eqnarray}}

\def\and{{\rm and\ }}
\def\be{\begin{equation}}
\def\ee{\end{equation}}

\slugcomment{Submitted to the Astrophysical Journal}

\shorttitle{Formation location of Uranus and Neptune}
\shortauthors{Kavelaars et al.}

\begin{document}

\title{On the formation location of Uranus and Neptune as constrained by dynamical and chemical models of comets}

\author{J.J. Kavelaars}
\affil{Herzberg Institute of Astrophysics, National Research Council of Canada, 5071 West Saanich Road, Victoria, BC V9E 2E7, Canada}
\email{JJ.Kavelaars@nrc.gc.ca}
\author{Olivier  Mousis}
\author{Jean-Marc Petit}
\affil{Institut UTINAM, CNRS-UMR 6213, Observatoire de Besan\c{c}on, BP 1615, 25010 Besan\c{c}on Cedex, France}
\author{Harold A. Weaver}
\affil{Space Department, Johns Hopkins University Applied Physics Laboratory, 11100 Johns Hopkins Road, Laurel, MD 20723-6099, USA}

\begin{abstract}
The D/H enrichment observed in Saturn's satellite Enceladus is remarkably similar to the values observed in the nearly-isotropic comets.
Given the predicted strong variation of D/H with heliocentric distance in the solar nebula, this observation links the primordial source region of the nearly-isotropic comets with the formation location of Enceladus.  That is, comets from the nearly-isotropic class were most likely fed into their current reservoir, the Oort cloud, from a source region near the formation location of Enceladus.
Dynamical simulations of the formation of the Oort cloud indicate that Uranus and Neptune are, primarily, responsible for the delivery of material into the Oort cloud.  In addition, Enceladus  formed from material that condensed from the solar nebula near the location at which Saturn captured its gas envelope, most likely at or near Saturn's current location in the solar system.
The coupling of these lines of evidence appears to require that Uranus and Neptune were, during the epoch of the formation of the Oort cloud, much closer to the current location of Saturn than they are currently.  
Such a configuration is consistent with the Nice model of the evolution of the outer solar system.
Further measurements of the D/H enrichment in comets,  particularly in ecliptic comets, will provide an excellent discriminator among various models of the formation of the outer solar system.
\end{abstract}


\section{Introduction}

Levison (1996), following on previous work by Carusi et al. (1987) and others, proposes two broad classes of comets, the ecliptic and the nearly isotropic.  Objects are selected into these dynamical classes by their Tisserand parameter with respect to Jupiter.  
Levison finds that the value $T_J \sim 2$ results in a secure boundary between comets from different reservoirs.  
Different reservoirs likely indicate different source regions within the primordial solar nebula.
Determining the source regions from which the comet reservoirs were first populated, and modeling the chemical evolution of those source regions as constrained by observations of comets,  will provide important clues on the physical and chemical structure of the primordial solar system.  

A comet's origins in the primitive nebula can be probed by examining the degree to which fossil deuterium
is enriched compared to the protosolar abundance. Calculations of the temporal and radial evolution of the deuterium enrichment in the solar nebula 
can reproduce existing D/H measures for comets (Drouart et al. 1999; Mousis et al. 2000; Horner et al. 2007). These calculations show that the deuterium enrichment in water ice strongly depends on the distance from the Sun at which the ice was formed. Comparing the D/H value measured in comets with those predicted by such models allows retrieval of their formation location. 

The measurement of the D/H ratio at Enceladus by the Ion and Neutral Mass Spectrometer aboard the $Cassini$ spacecraft (Waite et al. 2009) provides a new, and tighter, constraint on the deuterium enrichment profile in the outer solar nebula prompting us to reconsider models presented in previous works.   We pay particular attention, in this analysis,  to the source region of the reservoir of nearly-isotropic  comets under the conditions described in the Nice model scenario (Levison et al. 2008) of the formation of the outer solar system.   
We demonstrate that the measured D/H abundance ratios for Oort cloud comets are consistent with their formation having been in the 10-15 AU zone of the solar system.  Further, comets with (D/H)$_{\tt{H_2O}} \lesssim 5\times10^{-4}$  are precluded from forming more than $\sim 15$ AU from the Sun.

\section{Reservoirs of comets and their source regions}
 
The `cometary reservoir' is the region of semi-stable phase space from which comets are currently being delivered, while the `source regions' are those parts of the primitive nebula in which the comets formed and were then delivered to the reservoirs.  Ecliptic and isotropic comets are being delivered from at least two distinct reservoirs and, as such, are likely from different source regions. 

The reservoir of the ecliptic comets has been demonstrated to be the Kuiper belt and may be, more precisely, the `scattered disk' component of that population (Duncan \& Levison 1997).  
The source region of the Kuiper belt is a matter of current debate. In the Nice model,  Uranus and Neptune originate in the 10--15 AU region of the primordial solar system and later are transported to their current locations via dynamical interactions.  During this process, material in the 20--30 AU region is deposited into the Kuiper belt and scattered disk.  More classically, the source of the Kuiper belt may be the remnant of an in situ population.  
Regardless, the ecliptic comets now being delivered from some part of the Kuiper belt formed beyond the formation location of Neptune.   

For the isotropic comets the reservoir region is, generically, the Oort cloud (see Dones et al. 2004 for a good review).  Some fraction of the isotropic comets with $a < 20,000$~AU may arrive from the `innermost' component of this distribution (Kaib \& Quinn 2009), the remainder coming from the outer Oort cloud.  
Modelling of delivery into the Oort cloud reservoir (e.g., Dones et al. 2004) generally finds this process to be controlled by Uranus-Neptune scattering.
The discovery of objects with large peri-centres,  such as 2000 CR105 (Gladman et al. 2002) and (90377) Sedna (Brown, Trujillo \& Rabinowitz 2004), motivated Brasser, Duncan \& Levison (2006) and Kaib \& Quinn (2008) to examine the dynamics of Oort cloud formation in the presence of a stellar birth cluster.  They find that material from the Uranus-Neptune region of the primordial solar system is effectively transported into the inner and outer Oort cloud regions, the reservoirs of future nearly-isotropic comets.  Including the effect of gas-drag in the solar nebula (Brasser, Duncan \& Levison 2007) allows material in the `innermost' Oort cloud to also be delivered by Jupiter and Saturn.  Uranus and Neptune, however, dominate the post-nebula delivery.  Thus, the Uranus-Neptune region appears to be the likely source of material that now inhabits the inner and outer Oort clouds.  

If Uranus and Neptune originated at (roughly) 12 and 15 AU then material currently being delivered from the Oort cloud reservoir should have originated from a source much closer to the Sun than in cases where Uranus and Neptune formed at or near their current locations ($\sim$20 \& 30 AU).  A tracer of the chemical evolution of the primordial solar system that is sensitive to variations in the physical conditions between 10 and 30 AU, an example of which is described in the next section, provides a discriminator between these formation scenarios.

\section{Isotopic fractionation of deuterium in the solar nebula}

The main reservoir of deuterium in the solar nebula was molecular hydrogen (HD vs. H$_2$), and ion-molecule reactions in the interstellar medium (see e.g. Brown \& Millar 1989) causes fractionation amoung deuterated species. Consequently, in the pre-solar cloud, fractionation resulted in heavier molecules being enriched in deuterium.
As the second most abundant hydrogen bearer in the solar nebula, water became the second largest deuterium reservoir. 

We follow the approaches of Drouart et al. (1999) and Mousis et al. (2000) who described the evolution of the deuterium enrichment factor, $f$,  that results from the exchange between HD and HDO.  $f$ is defined as the ratio of D/H in H$_2$O to that in molecular H$_2$.  
Here we consider an additional constraint that tightens the deuterium enrichment profiles calculated in Mousis et al. (2000). 
The recent measurement by the $Cassini$ spacecraft of the D/H ratio in the plumes of Enceladus, one of the ice-rich regular moons of Saturn, shows that this value is in the same range as those measured in comets (see Table~\ref{DH}). 
If Enceladus formed near the current location of Saturn, (which likely formed within $\sim$1 AU of its current location),  we can then pin the value of $f$ at this location in the nebula

We use the equation of diffusion describing the evolution of $f$ and the solar nebula model depicted by Mousis et al. (2000) in which the disk viscously spreads out with time under the action of turbulence. The equation of diffusion takes into account the isotopic exchange between HDO and H$_2$ in the vapor phase, and turbulent diffusion throughout the solar nebula. The diffusion equation remains valid as long as H$_2$O does not condense, which implies that the value of $f$ is ``frozen'' into the microscopic ices present at the time and location of condensation. As the grains reach millimeter size, they begin to decouple from the gas leading to the formation of planetesimals. This implies that the enrichment factor $f$ acquired by planetesimals is that of the microscopic grains from which they formed, irrespective of the planetesimals subsequent evolution. We consider the case where the cometesimals (planetesimals that find their way to the cometary reservoirs) were accreted only from icy grains formed locally to the reservoir source.  This statement is consistent with Horner et al. (2007) who conclude that there is little diffusion due to turbulence with grain transport limited to only a few AU . This implies that the D/H ratio in the deuterated ices in comets is the value at the time and location at which they condensed and may be used to discriminate among models of the outer solar system's evolution.

Figure~\ref{prof} describes the evolution of $f$ as a function of distance from the Sun in the case of the solar nebula defined by the parameters $\alpha$ = 0.003, $R_{D0}$ = 15 AU and $M_{D0}$ = 0.06, each of them figuring within the range of possible values determined by Mousis et al. (2000). As in previous work, we assume that $f$ is constant at $t$ = 0 irrespective of the heliocentric distance and corresponds to the value measured in the highly enriched component found in LL3 meteorites (D/H~=~$(73~\pm~12)~\times~10^{-5}$; Deloule et al. 1998) compared to the protosolar value ($2.1 \pm 0.4 \times 10^{-5}$; Geiss \& Gloeckler 1998). The highly enriched component in LL3 meteorites is presumed to originate from ISM grains that were not reprocessed when entering the nebula (Mousis et al. 2000) and is consistent with D/H measurements from {\em Infrared Space Observatory} in grain mantles in W33A (Teixeira et al. 1999). 

For the adopted set of parameters, the deuterium enrichment profile simultaneously matches the nominal D/H value measured in H$_2$O in the moderately enriched component of LL3 meteorites at 3 AU and at the current heliocentric distance of Saturn matches the D/H enrichment of Enceladus.  We were unable, in this investigation, to find models matching both the moderately enriched component of the LL3 meteorites at 3 AU and the value at Enceladus at 10 AU that did not also require the value of $f$ to increase to much larger values in the region beyond 15 AU.  Thus, the result that $f$ in the 20-30 AU zone should have exceeded $\sim 25$ is a generic outcome of the temperature evolution of the disk, when constrained by the D/H measured at Enceladus, and not particularly dependent on the model of that evolution.

\section{Interpretation of the deuterium to hydrogen ratio measured at Enceladus by the $Cassini$ spacecraft}

One could argue that the building blocks of Enceladus were formed in Saturn's subnebula, implying that the D/H ratio in H$_2$O measured at this satellite by the $Cassini$ spacecraft might not be representative of the one acquired by planetesimals condensed in Saturn's feeding zone in the solar nebula. In order to show that this hypothesis is unlikely, we have performed calculations of the evolution of the D/H ratio in H$_2$O in Saturn's initially hot subnebula. The hypothesis of an initially hot subnebula is required if one wants to assume that the building blocks of the regular icy satellites, including Enceladus, were formed $in~situ$. 
To do so, we have used the same turbulent disk model utilized to describe the evolution of the D/H ratio in water in the solar nebula, but in a version scaled to the plausible size and properties of the Saturn's subnebula. This model has already been used to describe the thermodynamic evolution of cold subnebulae around Saturn and Jupiter (Mousis et al. 2002a; Mousis et al. 2002b; Mousis \& Gautier 2004). Here we consider the subdisk parameters of the initially hot Saturn's subnebula depicted by Alibert \& Mousis (2007) and whose evolution was constrained by Saturn's formation models. The viscosity parameter, the initial mass, and outer edge of our Saturn's subnebula have then been set to 2 $\times$ $10^{-4}$, 7 $\times$ $10^{3}$ Saturn's mass and 200 Saturnian radii, respectively. 

\begin{figure}
\resizebox{\hsize}{!}{\includegraphics[angle=0]{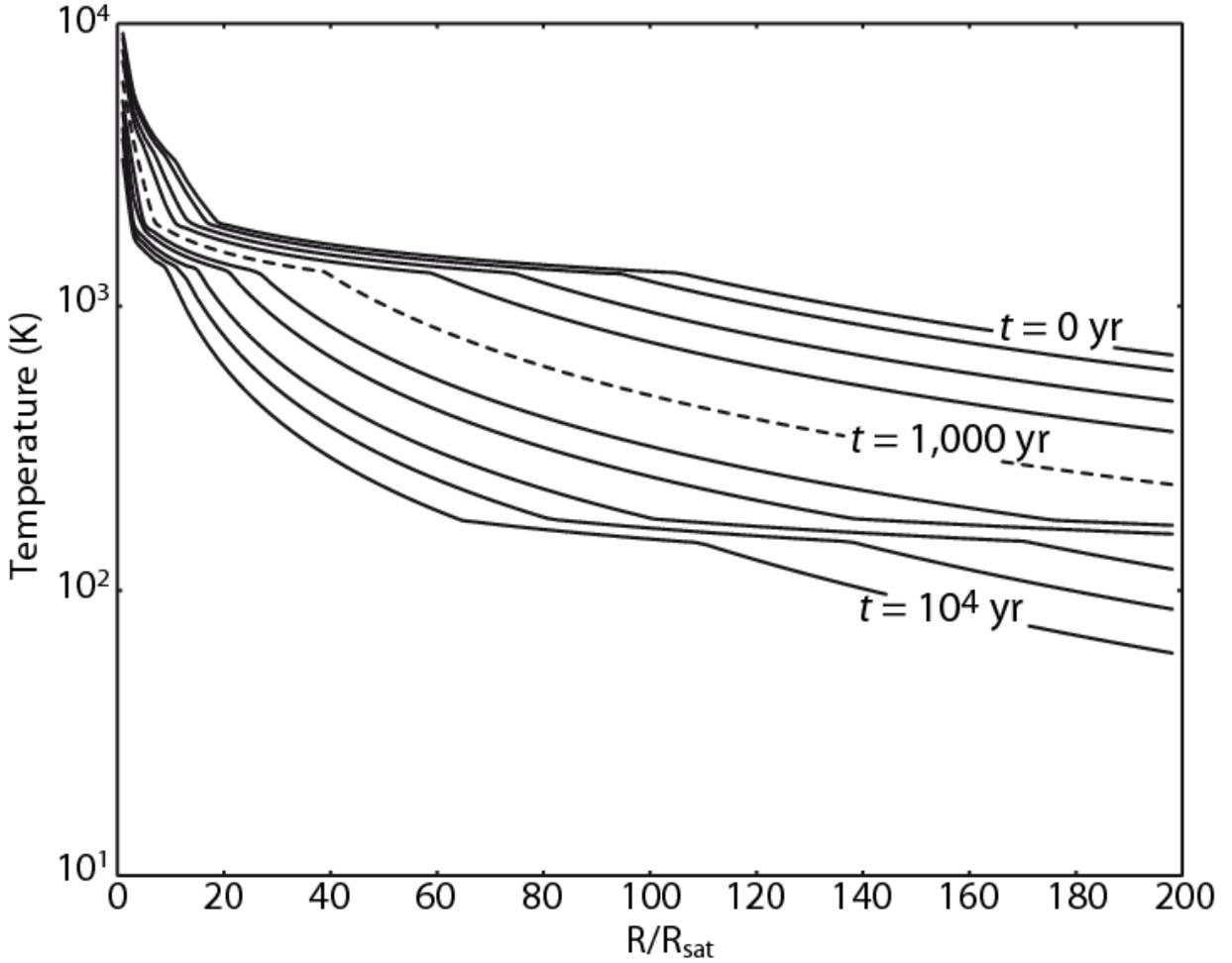}} \caption{ Temperature profiles at different epochs in the midplane of the Saturnian subnebula, at times (from top to bottom) $t$ = 0, 5, 200, 400, $10^{3}$, 2 $\times$ $10^{3}$, 3 $\times$ $10^{3}$, 5 $\times$ $10^{3}$, 7 $\times$ $10^{3}$, and  $10^{4}$ yr as a function of the distance from Saturn in units of Saturn radii. Dashed curve corresponds to the epoch $t = 10^{3}$ yr at which the deuterium enrichment factor of the D/H ratio in H$_2$O reaches the protosolar value in the whole subdisk (see Figure~\ref{profD}).} 
\label{profT}
\end{figure}

Figure~\ref{profT} shows the temporal evolution of the temperature profile in the midplane of Saturn's subnebula. 
Because the initial temperature of the subnebula is very high, any icy planetesimal entering the subdisk at early epochs of its evolution should be devolatilized and would then enrich the gas phase of the disk. In this model, ice forms again at the outer edge of the subnebula at $t$ $\sim$ 3 $\times$ $10^{3}$ yr (once the gas temperature has decreased down to $\sim$155 K at the corresponding pressure conditions) and its condensation front reaches the orbit of Enceladus after only a few dozen thousands of years of evolution.

\begin{figure}
\resizebox{\hsize}{!}{\includegraphics[angle=0]{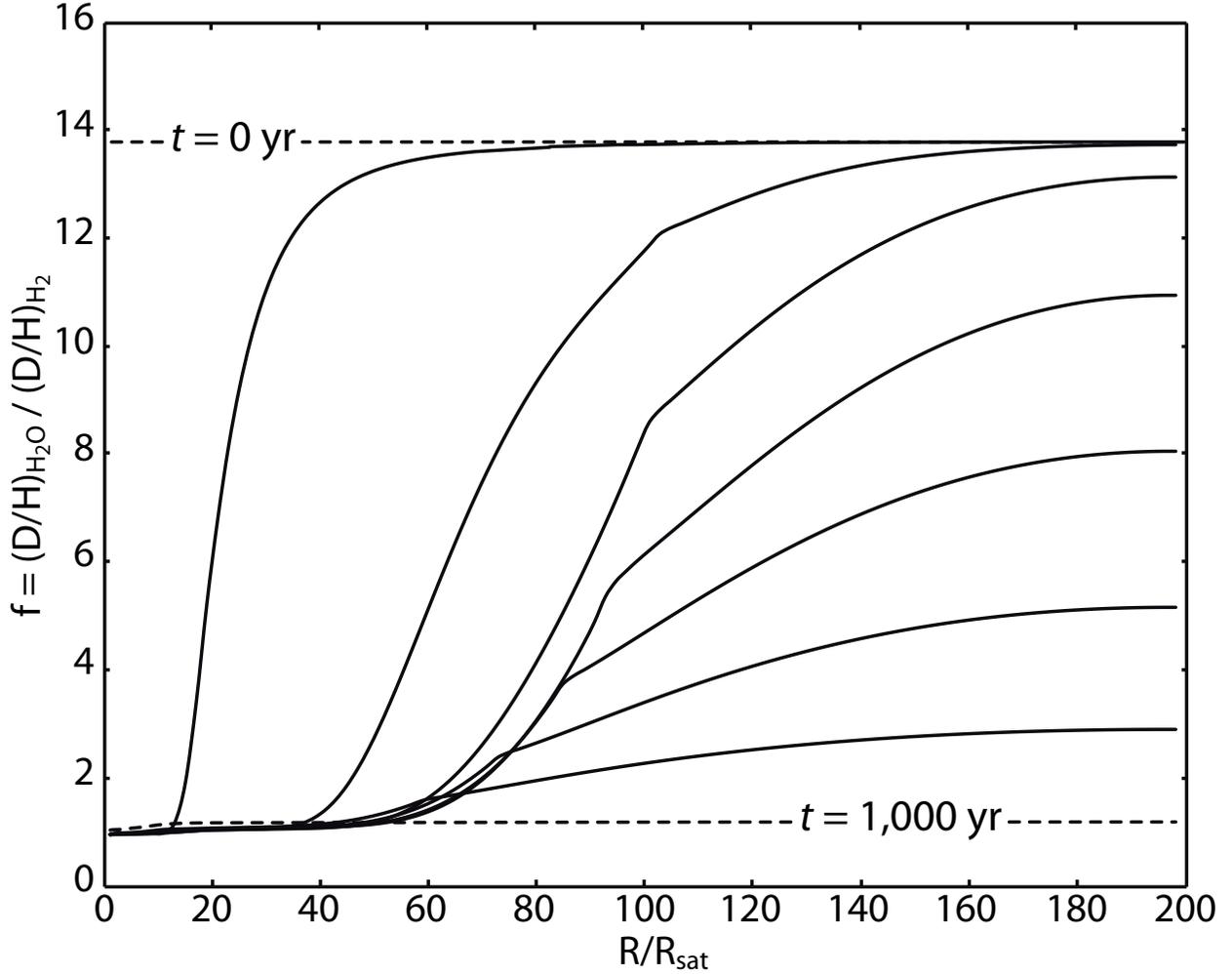}} \caption{ Enrichment factor $f$ of the D/H ratio in H$_2$O with respect to the protosolar value in the subnebula midplane, as a function of the distance to Saturn (in units of Saturn radii), at times (from top to bottom) $t$ = 0, 0.1, 5, 20, 50, 100, 200, 400,  and $10^{3}$ yr, see the text for details. 
The value for $f$ at $t$ = 0 is taken to be equal to 13.8 (the value measured at Enceladus by the $Cassini$ spacecraft), irrespective of the distance to Saturn in the subdisk. 
At the epoch $t = 10^{3}$ yr the deuterium enrichment factor in H$_2$O reaches the protosolar value in the whole subdisk. For Saturn D/H $= 1.7^{+0.75}_{-0.45} \times 10^{-5}$ (Lellouch et al., 2001) resulting in $f\sim0.8$.} 
\label{profD}
\end{figure}

Figure~\ref{profD} represents the evolution of the D/H ratio in H$_2$O in the subnebula described with the same approach as in Section 3. We have assumed that the deuterium enrichment factor, $f$, is equal to 13.8 (i.e., the value measured at Enceladus by the $Cassini$ spacecraft) in the whole subnebula at $t$ = 0. Due to the high temperature and pressure conditions that favor the isotopic exchange between H$_2$O and H$_2$ within the subnebula, $f$ rapidly diminishes and converges toward 1 in about 1000 years, prior to the condensation of ice (see dashed curve in Figure~\ref{profD}). We find that planetesimals should present D/H ratios in H$_2$O very close to the protosolar value if they were condensed within Saturn's subnebula. The isotopic exchange is so efficient at the temperature and pressure ranges likely to have been present the Saturn subnebula that $f$ would converge towards $\sim$ 1 for nearly any choice of initial value.  The $Cassini$ measurement at Enceladus shows that the D/H ratio in H$_2$O present in the plumes is strongly over-solar and we conclude that the building blocks of this satellite must have formed in the solar nebula.

\section{Implications for the primordial origin of comets}

The D/H ratio for cometary water ice is available for only a limited sample of comets, with two measurements available for only two.  These measurements (see Table~\ref{DH} and references therein) have been conducted using a variety of methods: remote UV spectroscopy (C/2001 Q4 (NEAT)),  mass spectroscopy (1P/Halley), radio spectroscopy (C/1996 B2 (Hyakutake) and C/1995 O1 (Hale-Bopp)), and infrared spectroscopy (8P/Tuttle).   Despite the variety of techniques used for the cometary measurements and the limitations of each (see the footnotes to Table 1), a remarkably narrow range of D/H values have been reported. Table 1 summarizes those results and also includes the result for the D/H ratio of Enceladus from $Cassini$.
The taxonomic classification using the system Levison (1996) is also provided.   All  of these comets are members of the nearly-isotropic class.
Comets like C/2004 Q4 (NEAT) are almost certainly to have originated from the outer Oort cloud reservoir while the `external' and `Halley type' comets may, in fact, come from the inner-most Oort cloud (Kaib \& Quinn 2009).

\subsection{Isotropic comets}

The isotropic comets have their origin in some part (inner-most, inner or outer) of the Oort cloud.  Based on the value of $f$ observed in the nearly-isotropic comets ($\sim 13 - 23$) and our modeling of the evolution of $f$, the cometesimals are most likely to have been delivered into the Oort cloud from a source region between 10 and 14 AU from the Sun.  We find that the value of $f$ interior to $\sim$10~AU is too low for the nearly-isotropic comets, implying that Jupiter and Saturn where not responsible for populating this reservoir.  Further, in the classical picture of solar system formation, where Uranus and Neptune form near their current locations of 20 and 30 AU, the ice-giants would have delivered cometesimals to the Oort cloud with values of $f > 25$, which is not seen.  
We find that, for our model of deuterium evolution, having a value of $f \sim 15$ (as required by the Enceladus measurement) at 10 AU  and $f \sim 15$  at 25 AU is not possible.

The Nice model for the formation of the solar system, however,  asserts that the formation  location of Uranus/Neptune, and presumably  then the region from which they delivered the majority of the material  into the Oort cloud, was considerably nearer to present day Saturn, between 11 and 13~AU for Uranus and 13.5 and 17~AU for Neptune (Tsiganis et al., 2005).  This is  precisely that zone of the primordial solar system which our modeling indicates cometesimals would have formed with values of $f$ similar to that observed in the nearly-isotropic comets.  Thus, the current measured values of  $f$ in the isotropic comet population appears to support a more compact configuration for the early solar system.  Our knowledge of the dynamics of the formation of the Oort cloud from a compact configuration remains uncertain, indeed the origin of the Oort cloud comets maybe varied (Clube \& Napier 1984, for example).   The homogeneity of D/H measures in Oort cloud comets and similarity of those values to that measured for Enceladus provides an interesting constraints for such scenarios.

\subsection{Ecliptic comets}

At present, no comets in the ecliptic class have known D/H levels.  The $Rosetta$ mission, currently en route to the ecliptic comet 67P/Churyumov-Gerasimenko may alter this situation.  Dynamical processes that populate the ecliptic comet reservoir (either the Kuiper belt, scattering disk, or some combination) all draw their source populations from beyond the orbit of Neptune (at least beyond 17 AU).  Based on our model of the radial dependence of $f$ (see Figure~\ref{prof}), we predict that the measured D/H ratio in the ecliptic comet population should exceed 24 times solar.

\section{Conclusions}

\begin{figure}
\resizebox{\hsize}{!}{\includegraphics[angle=-90]{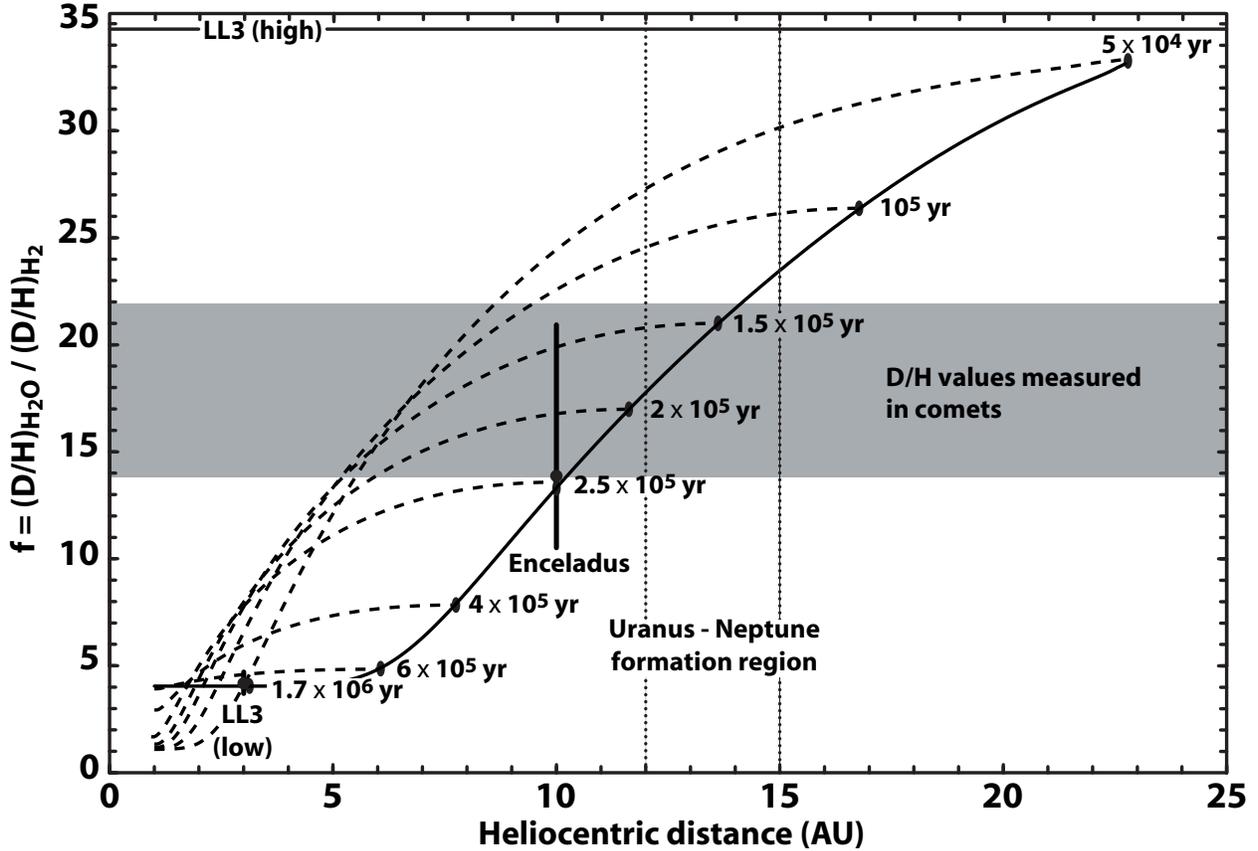}} \caption{Enrichment factor $f$ as a function of the heliocentric distance. The dashed curves correspond to the evolution of $f$ in the gas phase prior to condensation terminated by dots at the heliocentric distance where H$_2$O condenses at the given epoch. The solid curve represents the value of $f$ acquired by  ice as a function of its formation distance in the nebula. D/H enrichments in LL3(low and high) meteorites and Enceladus are shown for comparison.  We  take the LL3(high) value as the initial, protosolar, value.  The vertical dotted lines enclose the  source region of Uranus and Neptune in the Nice model. The gray area corresponds to the dispersion of the central values of the $f$ in the  comets for which measurements are available (see Table \ref{DH}). } 
\label{prof}
\end{figure}

1P/Halley, 8P/Tuttle, C/1995 O1 (Hale-Bopp), C/1996 B2 (Hyakutake) and C/2001 Q4 (NEAT) all have D/H values that are consistent with or slightly larger than that of Enceladus.  These comets are all members of the nearly-isotropic class and are, thus, drawn from a reservoir in some part of the Oort cloud.  Based on dynamical arguments, the Oort cloud itself was fed by material from the Uranus/Neptune region.  
Our modeling of the dependence of $f$  (pinned by the measured deuterium enrichment of Enceladus)  on formation location (see Figure~\ref{prof}) precludes these comets from having formed beyond $\sim$15~AU from Sun.
This implies that Uranus and Neptune were originally closer to the current location of Saturn than observed today, a configuration quite similar to that preferred in the Nice model.  Future space probe missions and improved  remote sensing capabilities will likely provide a larger number and variety of cometary D/H measurements and will surely increase the constraints on the primordial configuration from which the planetary system evolved to its current state.

\acknowledgments
Helpful advice provided by Ramon Brasser is gratefully acknowledged.  J. Kavelaars acknowledges support provided by Embassy France. O. Mousis acknowledges support provided by the Centre National d'Etudes Spatiales.

\onecolumn 

\begin{deluxetable}{llcll}
\rotate
\tabletypesize{\scriptsize}
\tablecaption{Deuterium measurements in H$_2$O in Enceladus and in different comets}
\tablenum{1}
\tablehead{\colhead{Name} &\colhead{(D/H)$_{\tt H_2O}$} ($\times~10^{-4}$) & \colhead{$f$\tablenotemark{a}}& \colhead{Reference} & \colhead{Object Class} }
\startdata
LL3 (high)                            &  $7.3~\ \pm~1.2       $ & 34.8 &  Deloule et al. (1998) & \\
LL3 (low)          		         &  $0.88~\pm~.11     $ & \ 4.2  & Deloule et al. (1998) & \\
Enceladus	       		&  $2.9^{+1.5}_{-0.7}$ & 13.8        & Waite et al. (2009)\tablenotemark{b} 			& Regular icy satellite of Saturn\\
C/2001 Q4 (NEAT)		&  $4.6 \pm 1.4          $ & 21.9 	& Weaver et al. (2008)\tablenotemark{c}			& Isotropic, new \\
1P/Halley  			&  $3.1^{+0.4}_{-0.5}$ & 14.7  	& Balsiger et al. (1995)\tablenotemark{d} 			& Isotropic, returning, Halley type\\
\nodata		        	         	&  $3.2~\ \pm~0.3$  	   & 15.0	& Eberhardt et al. (1995)\tablenotemark{e}		& \nodata \\
C/1996 B2 (Hyakutake) 	&  $2.9~\ \pm~1.0$  	   & 13.8	& Bockel{\' e}e-Morvan et al. (1998)\tablenotemark{f} 	& Isotropic, returning, external \\
C/1995 O1 (Hale-Bopp)	&  $3.3~\ \pm~0.8$	   & 15.7       	& Meier et al. (1998)\tablenotemark{f} 			& Isotropic, returning, external \\
\nodata                               	&  $4.7~\ \pm~1.1$        &  22.4      & Crovisier et al. (2004)\tablenotemark{f,g}                & \nodata\\
8P/Tuttle                             	&  $4.1~\ \pm~1.5$        & 19.5       & Villanueva et al. (2009)\tablenotemark{h}                              & Isotropic, returning, Halley type \\
\enddata

\tablenotetext{a}{Enhancement of D/H in H$_2$O compared to the protosolar D/H value  of $(0.21~\pm~0.05)\times 10^{-4}$  (Geiss \& Gloeckler 1998) }
\tablenotetext{b}{D/H in molecular hydrogen in the plume of material ejected from Enceladus, D/H in molecular hydrogen should be representative of D/H in water.}
\tablenotetext{c}{Ultraviolet measurements of atomic D and H in the coma, assumes HDO and H$_2$O photolysis are the exclusive sources of D and H.}
\tablenotetext{d}{Ion mass spectrometer measurements of D/H in the hydronium ion (H$_3$O$^+$), assumes same ratio holds in water.}
\tablenotetext{e}{Neutral and ion mass spectrometer measurements of D/H in the hydronium ion (H$_3$O$^+$), corrected for fractionation in the ratio for water.}
\tablenotetext{f}{HDO production rate derived from the measurement of a single submillimeter HDO line and a water production rate obtained from other observations made at a different time.}
\tablenotetext{g}{The authors also reported an upper limit of D/H$\lesssim 1.8~\times~10^{-4}$ using  a different line, which is inconsistent with their detections from two other HDO lines.}
\tablenotetext{h}{The listed D/H is consistent, at the 3$\sigma$ level, with D/H $< 4.35~\times~10^{-4}$. }

\label{DH}
\end{deluxetable}
\clearpage

\clearpage
\clearpage

\end{document}